\documentclass[preprint]{aastex}

\newcommand{\eqnref}[1]{(\ref{#1})}
\newcommand{\rmi}{{\rm i}}			
\newcommand{\rmd}{{\rm d}}			
\newcommand{\rme}{{\rm e}}			

\begin{document}

\title{Helioseismic detection of deep meridional flow}

\author{Douglas Gough}
\affil{Institute of Astronomy and Department of Applied Mathematics and
Theoretical Physics, University of Cambridge, Cambridge CB3~0HA, UK}
\email{douglas@ast.cam.ac.uk}

\author{Bradley W. Hindman}
\affil{JILA and Department of Astrophysical and Planetary Sciences,
University of Colorado, Boulder, CO~80309-0440, USA}


\begin{abstract} 
Steady meridional flow makes no first-order perturbation to the frequencies
of helioseismic normal modes. It does, however, Doppler shift the local
wavenumber, thereby distorting the eigenfunctions. For high-degree modes,
whose peaks in a power spectrum are blended into continuous ridges, the
effect of the distortion is to shift the locations of those ridges. From
this blended superposition of modes, one can isolate oppositely directed
wave components with the same local horizontal wavenumber and measure a
frequency difference which can be safely used to infer the subsurface
background flow. But such a procedure fails for the components of the
more-deeply-penetrating low-degree modes that are not blended into ridges.
Instead, one must analyze the spatial distortions explicitly. With a simple
toy model, we illustrate one method by which that might be accomplished by
measuring the spatial variation of the oscillation phase. We estimate that
by this procedure it might be possible to infer meridional flow deep in the
solar convection zone.
\end{abstract}



\section{Introduction}
\label{sec:introduction}

The character of the meridional flow in the solar convection zone is a
fundamental component of modern mean-field and flux-transport dynamo models
\citep[e.g.,][]{Charbonneau:2005, Dikpati:2006, Rempel:2006a, Rempel:2006b}.
In the upper convection zone, meridional flows influence how magnetic fields
are transported from low to high latitudes, and how the fields are subducted
and subsequently processed in deeper layers. In some models the flows at the
base of the convection zone are believed to establish the period of the activity
cycle through the equatorward advection of the active-region belts.

Many dynamo models assume a simple structure for the meridional circulation
with a single cell in each hemisphere. However, global simulations of the
solar convection zone often exhibit complex multicellular patterns
\citep[e.g.,][]{Brun:2002, Brun:2004} which lack the simple structure sometimes
posited. Even those simulations that are dominated by a single cell in each
hemisphere \citep{Miesch:2008} usually possess countercells in thin layers
at the top and the base of the convection zone. Furthermore, while these 3-D
simulations generate meridional flows with amplitudes of about 15 to 20 m s$^{-1}$
close to the surface, at greater depths the flow velocity is about 10 m s$^{-1}$,
quite unlike the far smaller values that have been conjectured from simplified
mass-conservation arguments. The difference arises from systematic coupling
between the radial flow and the meridional circulation which results from
turbulent entrainment and detrainment processes.

Observations are not yet able to winnow the multitude of theoretical and
numerical models successfully. The meridional circulation deep below the
surface is presently poorly constrained, owing to a distressing paucity of
reliable information. A variety of techniques---including direct Doppler
measurement \citep[e.g.,][]{LaBonte:1982, Hathaway:1996}, magnetic feature
tracking \citep{Komm:1993, Svanda:2007}, and local helioseismology
\citep[e.g.,][]{Giles:1997, Haber:2002, Basu:2003, Zhao:2004}---have had
great success as indicators of meridional flow in the upper layers of the
convection zone. The concensus is that, in the near-surface layers, the
meridional flow is largely poleward and remarkably constant with depth,
with amplitude roughly 20 m s$^{-1}$. However, the flow is quite variable
with longitude, with active regions being sites of local inflow
\citep{Haber:2002, Gonzalez-Hernandez:2008, Hindman:2009}. While
local-helioseismological analyses have determined in exquisite detail the
flows in the upper 15\% of the convection zone, the detection of meridional
flow in deeper layers has been elusive. A variety of attempts to measure
the deep meridional flow have been made using both time--distance procedures
\citep{Giles:2000, Duvall:2003} and spectral procedures that seek frequency
shifts between northward- and southward-propagating waves \citep{Braun:1998,
Krieger:2007, Mitra-Kraev:2007}. But, stymied by the weak signal and the
large systematic errors, the flow below a depth of roughly 30 Mm remains
elusive \cite[for a detailed discussion of the inherent difficulties
see][]{Duvall:2009}.

Helioseismology determines flows below the solar surface from measurements
of the difference in the Doppler shift between acoustic waves propagating
in opposite directions. This is fundamentally true whether the method being
used is a spectral technique (such as ring analysis or global-mode analysis)
or a time--distance procedure. In spectral techniques, the Doppler shift
has traditionally been inferred by assessing the temporal frequency shift
of the acoustic-wave dispersion relation. In the presence of a steady flow,
waves propagating in opposite directions suffer opposite wavenumber shifts
which are often interpreted as Doppler frequency shifts at a given wavenumber
in a power spectrum of the wave field. However, this Doppler-frequency procedure
is not universally applicable, and, as we shall see in subsequent sections,
the measurement of meridional circulations in the lower half of the convection
zone is just such an instance where it should not be applied blindly. In
fact, in this instance, ``traditional" techniques will fail, and inversions
of so-called frequency splittings to determine the flow field are prone to
misinterpretation.

In this paper we present a 1-D analog of the effects of meridional flow on
acoustic modes. We use this simple model to illustrate the pitfalls that may
occur when using spectral techniques to measure meridional circulation in the
lower reaches of the convection zone and in the tachocline. We attempt to dispel
several misconceptions that have recently arisen concerning the nature of the
acoustic spectra in the presence of meridional flow, and suggest a novel
approach by which progress might be made. Explicitly, in Section 2, we discuss
the changing nature of the acoustic-mode spectrum from low to high degree. In
Section 3, we present the 1-D analog model and the artificial data that the model
generates. These data are used in Section 4 to illustrate an analysis scheme
designed to measure the flow speed from helioseismic observations. In Section 5
we consider briefly how the model and the analysis scheme generalize to the Sun,
and we discuss implications of our results.

\section{Acoustic Modes, Meridional Flows, and the Doppler Effect}
\label{sec:Merid_Doppler}

The Sun's acoustic oscillations, called $p$ modes (or p modes), span many wavelengths,
ranging from waves that girdle the Sun with only a few complete horizontal wavelengths
to waves with a thousand or more wavelengths. The nature of the $p$-mode power spectrum
changes dramatically between the long-wavelength and short-wavelength regimes.  Modes
of long wavelength (equivalently, with low harmonic degree $l$) have long lifetimes
and are able to encircle the Sun many times before they are eventually damped.
The constituent waves interfere with themselves constructively, and the resulting power
spectrum is dominated by distinct resonances. The linewidths of such modes are much
less than the frequency spacing between modes of neighboring degree; therefore, the
acoustic power is concentrated about well-defined, discrete frequencies.

High-degree modes, on the other hand, live too short a time for them to travel
around the Sun before being damped. Such waves have broad linewidths and,
therefore, modes of nearby degree blend to form a ridge of continuous power. The
transition between resonant frequencies and continuous ridges occurs when the time
required by a sound wave to encircle the Sun is comparable to the mode's lifetime;
that is typically when $150 \lesssim l \lesssim 200$. Since waves of degree 20 to 40
are needed to sample the tachocline and the lower half of the convection zone, we
note that any procedure to measure the meridional circulation in these layers must
take into account that the waves of interest are clearly in the regime of discrete
resonances.

We consider only those flows that, in a suitable rotating frame of reference $\Sigma$,
vary slowly as the solar cycle advances. Thus, we may consider such flows to be steady
compared to the time required for an acoustic wave to traverse the Sun. This means that
even in the presence of such flows, the Sun's acoustic oscillations can be decomposed
into well-defined normal modes: at low degree, in the presence of a meridional flow
that is steady in $\Sigma$, the acoustic field is dominated by normal modes with unique,
discrete frequencies.  Any decomposition of this wave field into northward- or
southward-propagating wave components does not produce waves with temporal frequencies
that are shifted relative to each other, contrary to what many previous spectral techniques
tacitly assume. Instead, the two wave components must add to form a normal mode with a
single frequency; the northward and southward wave components have the same frequency
but oppositely Doppler-shifted wavenumbers. Instead of a meridional flow shifting power
in temporal frequency at constant wavenumber, it modifies the effective wavenumber,
which in a spherical-harmonic decomposition spreads the power amongst neighboring
degrees at the same frequency.

In the case of high-degree waves, once again a meridional flow shifts the acoustic power
in degree at constant frequency. However, since the power is distributed continuously
along a ridge, this Doppler shift manifests as a translation of the entire ridge in
spectral space, and the movement of the ridge can safely be interpreted as a frequency
shift at constant degree, despite the underlying mechanism being a wavenumber shift.
Local-helioseismological techniques such as ring analysis have already successfully exploited
this property to measure meridional circulation in the upper layers of the convection
zone; however, we emphasize again that measuring the circulation at greater depth requires
the use of waves of low to intermediate degree ($l < 150$) whose discrete spectrum must
be accounted for properly.

We have repeated this point to emphasize an important matter: namely, that in the low-degree
region of the solar acoustic oscillation spectrum, the Doppler effect caused by meridional
flow results in a {\sl spread of power across wavenumber that cannot be interpreted as a
frequency shift}. Therefore, previous attempts to measure meridional circulation in the
lower reaches of the convection zone that have relied on presumed frequency shifts
\citep[e.g.,][]{Krieger:2007, Mitra-Kraev:2007} have resulted in misinterpretation. The
nature of this misinterpretation will be discussed in detail in Section \ref{sec:Discussion}.


\section{A Simple One-Dimensional Analog of Latitudinal Wave Propagation}
\label{sec:AnalogModel}

To expose the salient physics of the Doppler effect, as it pertains to the
issue at hand, we consider here a simple one-dimensional model---a model
that we shall generalize to the Sun in a straightforward manner in 
Section~\ref{sec:Discussion}. The model incorporates one-dimensional
propagation of acoustic waves though a uniform fluid (in a pipe) with constant
sound speed $c$. In the undisturbed state, the fluid moves with a small
velocity $U$ in the $x$-direction, the flow speed being subsonic with Mach
number $M=U/c \ll 1$. Initially, we shall presume this flow to be uniform,
but later in the paper we shall permit it to vary spatially. A good
approximation to such a system could be realized in practice by injecting
fluid through small holes in the sides of the pipe near one end, and withdrawing
it at the other. We shall require that the ends of the pipe (at $x=0$ and $x=a$)
are impenetrable, so the wave motion must vanish at each boundary.

Consider first the disturbance to be undamped, comprised of waves that have
existed for all time. In this case, the wave velocity $v(x,t)$ can be represented
as a superposition of normal, longitudinal, acoustic modes,

\begin{equation}
	v(x,t) = \sum_{l=1}^\infty A_l \; v_l(x,t),
\end{equation}

\noindent where $t$ is time and $x$ is position along the pipe ($x\in [0,a]$).
Each mode is characterized by a complex amplitude $A_l$ and a velocity eigenfunction
$v_l(x,t)$ given by

\begin{equation}
	v_l(x,t) = \sin(k_l x)\cos(\omega_l \tau), \qquad l = 1,2,3, ... \;,
	\label{eqn:EigenFunc}
\end{equation}

\noindent where

\begin{eqnarray}
        \omega_l &=& k_l c \left(1-\frac{U^2}{c^2}\right),
\\
	\tau(x,t) &\equiv& t + \frac{k_l U}{\omega_l c} x.
\end{eqnarray}

\noindent In these expressions the wavenumber $k_l$ is quantitized by the
impenetrability of the boundaries:

\begin{equation}
	k_l = \frac{l \pi}{a} .
	\label{eqn:k}
\end{equation}

We linearize in the Mach number $M$:

\begin{eqnarray}
        \omega_l &=& k_l c \; \left[1 + {\rm O}(M^2)\right],
\\
	\tau(x,t) &=& t + \frac{Ux}{c^2} \; \left[1 + {\rm O}(M^2)\right],
\end{eqnarray}

\noindent and to this order we note that the eigenfrequencies $\omega_l$ are
unaffected by the background flow $U$.  This is a general property, independent
of the dimensionality; for small flow speeds, the modification of the eigenfrequency
can be ignored, as it is second order in the Mach number.  \cite{Roth:2008} have
estimated that the second-order frequency shift due to meridional circulation
in the Sun is perhaps no more than 0.1 $\mu$Hz, a number that is probably so
small as to be immeasurable, at least in the immediate future..

The form of the disturbance $v_l$---given by Equation~\eqnref{eqn:EigenFunc}---can
equivalently be expressed as a superposition of two counter-propagating waves
of equal amplitude that are oppositely Doppler shifted: 

\begin{equation}
	v_l(x,t)=\frac{1}{2} \sin\left[\omega_l\left(t + \frac{x}{c - U}\right)\right]
		-\frac{1}{2} \sin\left[\omega_l\left(t - \frac{x}{c + U}\right)\right] .
	\label{eqn:CounterWaves}
\end{equation}

\noindent Evidently, to measure $U$ one must either measure the spatial variation
of the (temporal) phase in the normal mode $v_l(x,t)$, which is expressed by the
spatially varying time-like coordinate $\tau$, or measure the difference $2U$
between the phase speeds of the two counter-propagating waves in
expression~\eqnref{eqn:CounterWaves}.
We discuss both possibilities in greater detail in Section~\ref{sec:SpatialPhase};
however, in the solar context, it is the latter procedure that has commonly been
adopted in the past, and has been accomplished either by correlating disturbances
at pairs of spatially separated points and measuring the time lag for maximum correlation
(time--distance helioseismology) or by projecting the disturbance onto propagating
waveforms and trying to measure the apparent difference between the dispersion
relations for oppositely directed waves (global-mode helioseismology and ring analysis).
Here we explicitly discuss the latter spectral techniques, but we point out that our
cautionary remarks are pertinent to the interpretation of time--distance analyses as
well.

Note that in Equation~\eqnref{eqn:CounterWaves} the Doppler shift from the flow
manifests as a shift $\Delta k_l=\pm k_l U/c$ in the wavenumber, not in the
frequency $\omega_l$. This comes about because, in a reference frame fixed with
the pipe, a normal mode oscillates at all points in space with a unique frequency
$\omega_l$. Finally, we comment that this simple model can be generalized to
accommodate a spatially varying velocity $U(x)$: provided $k_l^{-1}$ is much less than the
scale of variation of $U(x)$, each eigenfunction is given to first order by
Equation~\eqnref{eqn:EigenFunc}, but now with

\begin{equation}
	\tau(x,t) = t + \frac{k_l}{\omega_l c} \int_0^x U(x^\prime) \; \rmd x^\prime \, .
		\label{eqn:tau}
\end{equation}


\subsection{Stochastic Excitation and Damping}
\label{subsec:ExciteDamp}

In reality the waves are intrinsically damped and are continually being stochastically
excited. In the Sun, acoustic wave excitation is predominantly by granules, by events
that are individually localized in both space and time. Therefore, a more appropriate
model is a superposition of impulsively excited, damped waves of the form

\begin{equation}
	v(x,t) =  \sum_{j=1}^J \sum_{l=1}^\infty A_{l j} \, v_l(x,t-t_j) \, H(t-t_j) \, \rme^{-\eta_l(t-t_j)},
	\label{eqn:Damped}
\end{equation}

\noindent where the summation over $j$ is a summation over $J$ distinct excitation
events, occurring at times $t = t_j$; $H(t)$ is the Heaviside step function; and
the parameters $A_{l j}$ and $t_j$ are random variables with respect to the index $j$.  Because
the exciting events are strongly localized in space, the dependence of the expectation
of $A_{l j}$ on $l$ is smooth, and indeed if the damping rate $\eta_l$ were independent
of $l$, the dependence would be flat for our one-dimensional model.

We now consider two possibilities. The first is when the damping time $\eta_l^{-1}$
is short or comparable with the return sound-travel time $2a/c$ along the tube. In
that case, there is little interference of the wave with itself, and an alternative
representation of the motion is simply an ensemble of waves with continuously varying $k$:

\begin{equation}
	v(x,t)=\sum_{j=1}^J \int_{-\infty}^\infty A_j(k) \, \sin(k x)\,\cos\left[\omega (\tau-t_j)\right] \, H(t-t_j) \; \rme^{-\eta(k)(t-t_j)} \,\rmd k,
\end{equation}

\noindent in which $\omega(k)=|k|c+kU$. This is the realm of the mode ridges
appropriate for solar modes of high degree.

In our model the counterparts to the low-degree modes have $\eta_l^{-1} \gg 2a/c$,
and therefore they interfere with themselves to form resonant standing oscillations
whose frequencies are unaffected by $U$ to first order.  This condition is equivalent
to the requirement that the damping rate $\eta_l$ is much less than the frequency
spacing $\Delta\omega = \pi c/a$ between modes of neighboring degree. Therefore,
the power associated with each mode is well separated spectrally and the low-degree
modes are in the realm of discrete peaks.


\subsection{Artificial Data Sets in Frequency Space}
\label{subsec:PowSpec}

If we assume that the duration of the observations is very long, we may apply
a continuous Fourier transform in time to Equation~\eqnref{eqn:Damped}. The
temporal Fourier transform is given by

\begin{eqnarray}
	\tilde{v}(x,\omega) &=& \sum_{l=1}^\infty  Q_l(x,\omega) \sum_{j=1}^J A_{l j} \; \rme^{\rmi \omega t_j},
		\label{eqn:FTv}
\\
        Q_l(x,\omega) &=& \frac{\rmi}{2} \sin(k_l x) \Bigg\{
                \frac{(\omega+\omega_l)-\rmi\eta_l}{(\omega+\omega_l)^2 + \eta_l^2}
                        \; \exp\left(\rmi\,\frac{k_l}{c} \int_0^x U(x^\prime) \, \rmd x^\prime\right)
		     	\nonumber
\\
        	&& \qquad \qquad \qquad + \frac{(\omega-\omega_l)-\rmi\eta_l}{(\omega-\omega_l)^2 + \eta_l^2}
                	\; \exp\left(-\rmi\,\frac{k_l}{c} \int_0^x U(x^\prime) \, \rmd x^\prime\right) \Bigg\} ,       
		\label{eqn:Q}
\end{eqnarray}

\noindent where $Q_l(x,\omega)$ is the temporal Fourier transform of
$v_l(x,t) \, H(t) \, \rme^{-\eta_l t}$, namely

\begin{equation}
        Q_l(x,\omega) = \int_{-\infty}^\infty \, v_l(x,t) \, H(t) \, \rme^{-\eta_l t} \rme^{\rmi \omega t}\;\rmd t.
\end{equation}

\noindent  In the preceding expressions, we have adopted a flow $U(x)$ that varies
slowly in space; hence $\tau$ is given by Equation~\eqnref{eqn:tau}. By way of
illustration, we choose $U(x)$ to be the fundamental sinusoid of amplitude $U_0$
that vanishes at each endpoint:

\begin{equation}
     U(x) = U_0 \sin(\pi x/a) .
	\label{eqn:Usine}
\end{equation}

\noindent For this special case, the integrals over $x^\prime$ appearing in
Equation~\eqnref{eqn:Q} can be evaluated analytically.

Expression~\eqnref{eqn:FTv} can be further simplified if we assume that the times
$t_j$ of excitation are uniformly distributed on the scale of the duration of the
observations, and that the number of such events $J$ is large. This permits the
summation over $j$ to be treated as a 2-D random walk in complex amplitude space.
Under these assumptions, Equation~\eqnref{eqn:FTv} can be expressed as follows:

\begin{equation}
	\tilde{v}(x,\omega) \approx J^{1/2} \sum_{l=1}^\infty {\cal A}_l \, \rme^{\rmi\theta_l} \, Q_l(x,\omega) ,
		\label{eqn:RandWalkDamped}
\end{equation}

\noindent where $\theta_l$ is a random phase with uniform distribution and ${\cal A}_l$
is the rms average over $j$ of the amplitudes $A_{lj}$.  Figure~\ref{fig:PowSpec0.1}
displays the power spectrum that results from applying a cosine window function and
taking the spatial Fourier transform of Equation~\eqnref{eqn:RandWalkDamped} with
$U_0$ chosen such that $M = U_0/c = 0.1$. While this value is unrealistically large
for meridional flows in the Sun, it was selected here purely for clarity of illustration.
The spatial Fourier transform of a function $f(x)$ is defined as follows:

\begin{equation}
	\tilde{f}(k) = \int_{-\infty}^\infty f(x) \, \rme^{-\rmi kx}\;  \rmd x,
\end{equation}

\noindent such that positive $k$ corresponds to waves propagating in the positive
$x$-direction.

The power in each normal mode is confined in a narrow frequency band about
$\omega = \omega_l$, and is further concentrated in wavenumber bands into two separate
branches, a positive-$k$ branch and negative-$k$ branch. These branches arise from
the decomposition of the normal mode into waves propagating in opposite directions.
The power in one branch is not necessarily symmetric with the other branch, since
the modes are capable of interfering with each other because of their finite lifetimes.
If the damping rates become sufficiently small compared to the frequency spacing between
modes ($\eta_l \ll \pi c/a$), this interference effect disappears. For all illustrations
appearing in this paper we have chosen a damping rate such that the ratio of the damping
rate to the frequency spacing is roughly 10 times larger than the actual value for
low-degree $p$ modes, but comparable to that for modes of intermediate degree. The
chosen value ($\eta_l/\Delta\omega = 2.5 \times 10^{-2}$) produces mode interference
with a small but noticeable effect.

Each mode appears with a series of spatial sidelobes. The amount of power in each
sidelobe is a complicated function of the spatial windowing and the deviation of the
eigenfunction from its zero-order (unperturbed by $U$) sinusoidal form. To generate
the power spectra, we applied a cosine window function to reduce the spread of
power due to windowing. Had we failed to apodize, the spread would have been worse;
a Fourier transform over a finite domain effectively has a top-hat window function
which spreads power rather broadly, due to its jump discontinuities. In power spectra
of solar oscillations, similar spatial windowing effects (often called spatial leaks)
arise automatically from a variety of sources, including the reduced spatial resolution
at the poles, the lack of sensitivity to motions transverse to the line of sight,
and the visibility of only one side of the Sun at any instant.


\subsection{Granulation Noise}
\label{subsec:Noise}

The Sun's meridional flow does not produce a Doppler shift as extreme as that
illustrated in Figure~\ref{fig:PowSpec0.1}, as the Mach number of the flow is
likely to be inordinately small, perhaps on the order of $10^{-4}$ or $10^{-5}$.
Whether such a small signal can be extracted from the observations depends on
the level of the noise. Therefore, in order to assess the detectability of a
particular meridional flow, we must include noise. For instruments such as MDI
on SOHO and HMI on the soon-to-be-launched SDO that use a photospheric line,
the primary source of noise is solar granulation. Owing to the relatively short
lifetime and small spatial scale of solar granulation, the noise spectrum generated
by granulation is broad both in frequency and wavenumber. Thus, in our 1-D analog,
we add white Gaussian noise, $N(x,\omega)$, to the wave field with a specified
rms amplitude:

\begin{equation}
	\tilde{v}(x,\omega) = N(x,\omega) + J^{1/2} \sum_{l=1}^\infty {\cal A}_l \, \rme^{\rmi\theta_l} \, Q_l(x,\omega) .
	\label{eqn:vwithNoise}
\end{equation}

The actual noise level for the MDI instrument is difficult to estimate because
the power that appears between the $p$-mode peaks is largely the contribution
of the overlapping wings of a large number of modes (personal communication,
J. Schou, August 2009). Since a direct measurement
of the level of noise for MDI is unavailable, we have attempted to estimate the
noise by scaling observations obtained with other instruments for which a careful
calibration of noise and signal has been obtained. Our estimate is based on the
level of noise measured in a 496-day data set from the GOLF instrument
\citep{Fletcher:2009}. This level is scaled to that pertaining to an equivalent
data set spanning 60 days, obtained from longitudinally averaged medium-$l$ MDI
data with 130 latitudinal pixels. Further, we have assumed that 10-years worth
of independent 60-day observations have been averaged together to reduce the noise.
The resulting signal-to-noise ratio is roughly 50.  A similar estimate can be
obtained by evaluating at 3 mHz the spectral power distribution of granular power
modeled by \cite{Harvey:1993} with a granular growth timescale of 220 s and rms
velocity of 0.6 km s$^{-1}$ \citep{Keil:1980}. The mode amplitudes were estimated
from measurements of the peak, spectral power density for low-degree modes obtained
by BiSON \citep{Chaplin:2005}. The scaling follows \cite{Christensen-Dalsgaard:1982}
and takes into account that the BiSON spectral line (K $\lambda769$ nm) is formed
in the chromosphere and that the MDI spectral line (Ni I $\lambda676.8$ nm)
is photospheric. Figure~\ref{fig:NoisySpec0.001} shows the power spectrum that
results when such noise is present for a flow with a smaller Mach number:
$U_0/c = 10^{-3}$. The Doppler shift is now too small to be seen by eye, and the
spatial sidelobes have diminished in prominence.

As mentioned previously, for normal modes, the Doppler shift acts on the local
wavenumber $k$ instead of acting on the mode frequency $\omega$, thus causing a
shift in the sidelobe pattern. This is clearly visible in Figure~\ref{fig:PowSpec0.1}$b$.
The prograde branch ($k>0$) possesses a steeper slope than the retrograde branch
($k<0$). Yet the mode power within each branch is centered around the same frequency.
Note that any attempt to measure a frequency shift by fitting a smooth profile
$F(\omega)$ (such as a Lorentzian) at fixed $k$ will be frustrated by the fact
that the power is confined to discrete frequencies that are unshifted (to first
order in the Mach number) by the velocity feld. This is illustrated further
in Figure~\ref{fig:Cuts}, where we present two pairs of cuts through the power
spectrum, one pair at constant wavenumber, and the other at constant frequency.
Each pair is composed of a cut through the retrograde branch and a separate
cut through the prograde branch.  The locations of these cuts are indicated in
Figure~\ref{fig:NoisySpec0.001}. The retrograde and prograde branches are shown
in red and blue, respectively, in Figure~\ref{fig:Cuts}.

Figure~\ref{fig:Cuts}$a$ depicts cuts through the spectrum as a function of frequency
for the constant value of the wavenumber $k$ corresponding to $l = 10$ ($k=10\pi/a$).
Rising up above the noise we see not only the mode frequency associated with that
value of $l$, but also all the frequencies of the modes with nearby values of $l$
(due to their spatial sidelobes and finite lifetime). This comb of frequencies has
an envelope which is modulated by the flow velocity, because both the principal peak
and the windowing sidelobes are modified by the Doppler effect. The envelope is by
no means Lorentzian, nor is it symmetric. Its exact shape is a complicated function
of the flow velocity $U(x)$, the spatial window
function, and the amplitudes and phases of the nearby modes. The Doppler effect is
more evident in cuts through the power spectrum at constant frequency, illustrated in
Figure~\ref{fig:Cuts}$b$. There is a small but clear systematic shift between the two
curves which is produced by the presence of the flow. The signature of the noise
appears to be much less prominent in Figure~\ref{fig:Cuts}$b$ than in
Figure~\ref{fig:Cuts}$a$. This is an illusion arising from the limited spectral
resolution in Figure~\ref{fig:Cuts}$b$. The curves appear smoother simply because
the noise (and signal) exist only in the finite domain $x\in[0,a]$; therefore, the
$k$-resolution $\delta k$ in the power spectrum is broad relative to the scale of
the plot: $\delta k \approx 2\pi/a$ (or 2 in the dimensionless units employed in
the figure). If a discrete Fourier transform were employed instead, only every other
tick mark would be sampled, as is indicated by the diamonds in Figure~\ref{fig:Cuts}$b$.
The noise's influence manifests predominantly as the growing disagreement between
the power in the sidelobes with increasing distance from the central peak.


\section{Measuring the Meridional Flow through Variations in the Spatial Phase}
\label{sec:SpatialPhase}

Since power in the spatial sidelobes is modified by a meridional flow, one might
attempt to measure the flow by a careful analysis of those sidelobes. Since the
sidelobes are sensitive to a large number of well-known, but poorly determined,
leakage effects (incomplete observational coverage of the solar surface, foreshortening,
velocity projection onto the line of sight, camera imperfections, etc.), we cannot
expect to measure the meridional flow from their absolute height. The only hope
that persists is
in attempting to measure the difference in power between the prograde and retrograde
branches. However, due to the fact that the power in the sidelobes is several orders
of magnitude lower than the central peak, such a measurement scheme is likely to
be fraught with signal-to-noise problems, particularly since the effect of the
meridional flow is expected to be rather weak anyway. Moreover, this scheme would
also be complicated by the fact that the complete pattern of the sidelobe structure
will not be available in finitely sampled data. (In Figure~\ref{fig:Cuts}$b$, the
diamond symbols indicate which wavenumbers would be represented by a discrete transform).

Because of these difficulties, we expect that a more promising means of measuring
the meridional flow seismologically would involve exploitation of the wave phase instead
of just its power. One needs to be careful here with the use of the word phase.
Does one mean the phase of the spatial and temporal transform, i.e., the phase of
the spectral component? Or does one mean the phase of the wave in configuration
space? In the remainder of this section, we propose a technique which is a hybrid
of these two options. To be explicit, the procedure derives the meridional flow
from measuring the phase of the wave expressed as a function of the temporal frequency
and of latitude.


\subsection{Spatial Variation of the Phase}
\label{subsec:PhaseModulation}

Examination of Equations~\eqnref{eqn:vwithNoise} and \eqnref{eqn:Q} reveals
that near a resonant frequency ($\left|\omega-\omega_l\right| \lesssim \eta_l$)
the waveform simplifies greatly:

\begin{equation}
	\tilde{v}(x,\omega_l) \approx N(x,\omega_l) + \frac{1}{2} J^{1/2} {\cal A}_l \, \eta_l^{-1}\, \rme^{\rmi \theta_l}
        	\, \sin(k_l x) \; \exp\left(-\rmi\frac{k_l}{c} \int_0^x U \, \rmd x^\prime\right).
\end{equation}

\noindent In the absence of noise, the phase $\phi_l$ of the wave is therefore
easily determined:

\begin{equation}
	\phi(x,\omega_l) \equiv \arg\left\{\tilde{v}(x,\omega_l)\right\} \approx \theta_l - \frac{k_l}{c} \int_0^x U \, \rmd x^\prime .
	\label{eqn:ResonantPhase}
\end{equation}

From Equation~\eqnref{eqn:ResonantPhase} it should be noticed that the modulation
of the phase, considered as a function of position $x$, has a relatively simple
dependence on the flow speed. In fact, the derivative of the phase with respect
to position depends only on the flow and several presumably known constants:

\begin{equation}
	\frac{\partial \phi(x,\omega_l)}{\partial x} = -\frac{\pi l}{ca} \, U(x) .
	\label{eqn:dphidx}
\end{equation}

\noindent Here we have utilized the quantization condition for the wavenumber,
Equation~\eqnref{eqn:k}, to obtain the dependence on the degree $l$.

Equation~\eqnref{eqn:dphidx} suggests an analysis scheme to extract the meridional
flow profile $U(x)$ from observations of the wave field. In the absence of noise,
one could sample the temporal Fourier transform of the wave field at a resonant
frequency $\omega = \omega_l$, compute the phase $\phi_l(x)$, differentiate with
respect to $x$, and finally use Equation~\eqnref{eqn:dphidx} to obtain the flow
as a function of $x$ (the analog of latitude in our 1-D model). Of course, in real
observations, noise will be an issue; however, the effects of noise can be reduced
by averaging the Fourier transform of the wave field over a band of frequencies
centered on a resonant frequency $\omega_l$ with a width of $2\eta_l$ to generate
an ``average mode eigenfunction" $\bar{v}_l(x)$. Since the phase of the noise is
uncorrelated over this frequency band, whereas the phase of the mode is correlated,
$\bar{v}_l(x)$ is less contaminated by noise than the wave signal at a single
frequency. Thus, a better estimate of the phase $\phi_l(x)$---and the amplitude
$V_l(x)$---of the waveform can be obtained through the following equations:

\begin{eqnarray}
	\bar{v}_l(x) &\equiv& \frac{1}{2\eta_l} \int_{\omega_l-\eta_l}^{\omega_l+\eta_l} \tilde{v}(x,\omega) \; \rmd \omega \, ,
		\label{eqn:AveEigfunc}
\\
	\phi_l(x) &\equiv& \arg\left\{\bar{v}_l(x)\right\} \, ,
		\label{eqn:Phase}
\\
	V_l(x) &\equiv& \left|\bar{v}_l(x)\right| \, .
		\label{eqn:Amplitude}
\end{eqnarray}

\noindent Why the amplitude $V_l$ is important will become apparent in a moment.

The amplitude and phase defined in this fashion have been computed for a variety
of model parameters. Figures~\ref{fig:CubicFlowPow} and \ref{fig:PhaseAmps} show
the results for a flow profile that is antisymmetric about $x = a/2$,

\begin{equation}
	U(x) = -12 \sqrt{3} \;\, U_0 \; a^{-3} \, x \, (x-a/2) \, (x-a) .
	\label{eqn:UCubic}
\end{equation}

\noindent This cubic form produces a flow speed that vanishes at both ends ($x=0$
and $x=a$) as well as in the middle $x=a/2$. The flow attains maximum and minimum
values of $\pm U_0$, and for Figures~\ref{fig:CubicFlowPow} and \ref{fig:PhaseAmps}
we have employed a large Mach number ($U_0/c = 0.1$) for clarity of illustration.
The flow is everywhere directed away from the center of the domain (i.e., the equator),
much as the Sun's meridional circulation is generally poleward in the outer
layers of the convection zone. We have intentionally chosen a functional form that
is non-sinudoidal to avoid using a flow that can be constructed from a finite number
of mode eigenfunctions.

Figure~\ref{fig:CubicFlowPow} shows the power spectrum obtained for this flow profile.
We point out that, unlike the previous example, the flow profile is antisymmetric,
and therefore, the mode power in each branch is Doppler broadened instead of Doppler
shifted.
Figure~\ref{fig:PhaseAmps} presents the square of the amplitude $V_l^2$ and the phase
$\phi_l$, both calculated from the same wave field for a selection of modes. The
square of the amplitude is scaled such that it would have unit maximum in the absence
of both the noise and the flow $U$. The phase is measured relative to the average phase
obtained by integrating the phase over the entire $x$-domain. 
The noise is clearly evident in both the amplitudes and the phases. Furthermore, the
phase is poorly determined wherever the amplitude is small. Thus, locations of poor
phase determination appear more frequently for higher-degree modes. The lack of
uniformity in the height of the amplitude peaks arises from both the noise and the
interference with nearby modes. We note further that the magnitude of the maximum phase
grows linearly with $l$, as is predicted by Equation~\eqnref{eqn:dphidx}.

The linear dependence on $l$ suggests a simple scheme to reduce the influence of the
noise further. An obvious procedure is to average $\phi_l$ over modes. But instead,
we average $\phi_l/l$ in order to remove the leading behavior on mode degree. Thus,
we define a mean phase function $\Phi(x)$ according to:

\begin{equation}
	\Phi(x) \equiv \sum_{l} w_l(x) \; \phi_l(x)/l,
	\label{eqn:AvePhase}
\end{equation}

\noindent for a set of weighting functions $w_l(x)$. The fact that the phase becomes
poorly determined when the amplitude $V_l(x)$ is small suggests that a weighting
based on the computed amplitudes would be effective. Since the amplitudes defined
by Equation~\eqnref{eqn:Amplitude} are non-negative, we choose to weight linearly
with $V_l$:

\begin{equation}
	w_l(x) = \frac{V_l(x)}{\sum_{l} V_l(x)} .
\end{equation}

\noindent From the phase function $\Phi(x)$ one can derive the flow velocity by
differentiation:

\begin{equation}
	c^{-1} U(x) \approx -\frac{a}{\pi} \frac{\rmd\Phi}{\rmd x} .
	\label{eqn:UfromPhase}
\end{equation}

The results of this procedure are illustrated in
Figures~\ref{fig:AvePhase0.1}--\ref{fig:AvePhase0.001}, which display the mean phase
function $\Phi(x)$ and the derived flow profile $U(x)$. The averaging has been perfomed
over all modes with $l\leq20$. Each figure corresponds to
a different maximum flow speed $U_0/c$, ranging from $10^{-1}$ to $10^{-3}$. The upper
panels show the mean phase functions and low-degree polynomial fits to the same. The
lower panels show the derived flow profiles obtained by differentiating the
{\sl polynomial fit} to the mean phase function. For large Mach numbers
($M\gtrsim10^{-2}$) the flow profile is recovered with great fidelity over the entire
spatial domain. As the Mach number decreases, the recovered flow profile begins to
diverge from the input flow profile, although initially only near the edges of the
domain. This property is to be expected: the edges of the domain are zeros of the
eigenfunctions for all degrees $l$, and, therefore, the determination of the phase
$\phi_l$ is poor there for all modes. As the Mach number falls further
($M\lesssim10^{-4}$), the noise comes to dominate and the recovered flow profile bears
little resemblance to the input values.


\section{Discussion}
\label{sec:Discussion}

The principles we have illustrated with our 1-D model can be applied to the
Sun. The most significant difference is that, for each mode, $c^{-1}U$ in
Equation~\eqnref{eqn:UfromPhase} should be replaced by its appropriately
weighted vertical average $\left<c^{-1}U_{ln}\right>$. For deeply penetrating modes,
the high-order asymptotic approximation to the oscillation eigenfunctions
provides an adequate first guide \citep{Gough:1993}, yielding a weighting
function $W(r,\omega/L)$ that changes form at the lower turning point
$r_{\rm t}$ given implicitly by $c(r_{\rm t})/r_{\rm t}=\omega/L$ where
$L = l + 1/2$. Below the lower turning point the weighting function vanishes,
$W = 0$, and above the lower turning point it is given by

\begin{equation}
	W\left(r,\frac{\omega}{L}\right) \sim \left(1-\frac{L^2c^2}{\omega^2 r^2}\right)^{-1/2}.
\end{equation}

\noindent Approximate weighting functions for higher-order modes are displayed
by \cite{Gough:1983}.

The analogous quantity to the return travel time in the one-dimensional model,
$2a/c$, is the circumundulation time $T_{ln} \sim 4 l\pi/\omega_{ln}$, the time
an acoustic-wave packet takes to traverse a great circle around the Sun. If the
damping time is short compared with the circumundulation time, $\eta_{ln} T_{ln} \gg 1$
(as is true for high-degree modes), the wave suffers little self interference,
the mode power blends with nearby modes forming a continuous ridge, and measuring
the apparent frequency shift $\delta\omega$ at given $l$ works well. But once
$\eta_{ln} T_{ln} \ll 1$ (low-degree modes), resonances dominate the wave field,
and the modes are well separated in the power spectrum, each with a frequency that
is unshifted by the meridional circulation.  Attempting to measure a shift by fitting
a Lorentzian to the frequency profile, or cross-correlating the power from poleward-
and equatorward-propagating waves, cannot therefore result in a meaningful nonzero
value. Instead, a background flow shifts and broadens the distribution of power
in degree $l$. As our discussion indicates, one expects a null frequency shift
for waves of low $l$: once $\eta_{ln} T_{ln} \lesssim 1$, the apparent shift
$\delta\omega$ must decline with decreasing $l$. Indeed that is just what appears
to have been observed \citep{Braun:1998, Mitra-Kraev:2007}, as is illustrated
in Figure~\ref{fig:MitraKraev}.

Misinterpreting this decline as a true decline in frequency has led to an erroneous
inference about the depth of the reversal of the Sun's meridional circulation. 
The circulation is well known to be directed poleward near the surface, and if
the radial average $\left<c^{-1}U_{ln}\right>$ over most of the convection zone (as
sampled by low-$l$ modes) vanishes, then the poleward surface flow must be
counterbalanced by a deep equatorward flow within the convection zone. If
$\delta\omega$ were a measure of that radial average, the flow reversal would be
located at the lower turning point of the modes near where $\delta\omega$ starts
to decline with decreasing $l$. That is just where the continuous ridges in the
power spectrum of the modes give way to distinguishable discrete peaks. The
transition occurs at $\nu/l \lesssim 20 \mu$Hz (see Figure~\ref{fig:MitraKraev})
which \cite{Mitra-Kraev:2007} acknowledge corresponds to a turning point at
about 40 Mm beneath photosphere. However, we have argued that this depth has
actually little, if anything, to do with the location of the reversal of the
real flow.

Since the eigenfrequencies are insensitive to the meridional flow, the only 
unambiguous way to detect meridional flow in the lower half of the convection
zone is via the structure of the oscillation eigenfunctions.
One telling property of that structure is that separability in space and time
is destroyed by the flow, causing latitudinal variation in temporal phase.
We propose here that the structure of the flow can in principle be determined
from direct measurements of that phase. We recognize that the phase variation
can equivalently be interpreted as a spatial variation at fixed time, which
together with an accompanying amplitude variation, distorts the eigenfunction
from its flow-free, spherical-harmonic counterpart. In a decomposition
into spherical harmonics, this distortion appears as a leakage of
power across degree $l$, that could conceivably be measured.  However, we
suspect that any attempt to measure that distortion directly is likely to be
much more susceptible to noise; it would also need to be separated from the
much larger contributions to the total distortion resulting from shear in the
zonal flow.

Finally, it behooves us to address the likelihood that a reversal in the flow
direction with depth will actually be detectable. Our simulations with 20 modes
indicates that, if our error estimates are realistic, the lower limit on the
Mach number for a measurable flow is between $10^{-4}$ and $10^{-3}$. That
corresponds to a flow velocity between about 20 and 200 m s$^{-1}$ at a depth
of about 150 Mm. The lower value is comparable with the flow speeds in the
convection-zone simulations by \cite{Miesch:2008}. So perhaps the detection of the flow reversal
in the Sun is almost in sight.


\acknowledgments

We acknowledge support from NASA through grants NNG05GM83G,
NNX08AJ08G, and NNX08AQ28G.




\def\etal{et al.}





\def\figone{%
\begin{figure*}%
        \epsscale{1.0}%
        \plotone{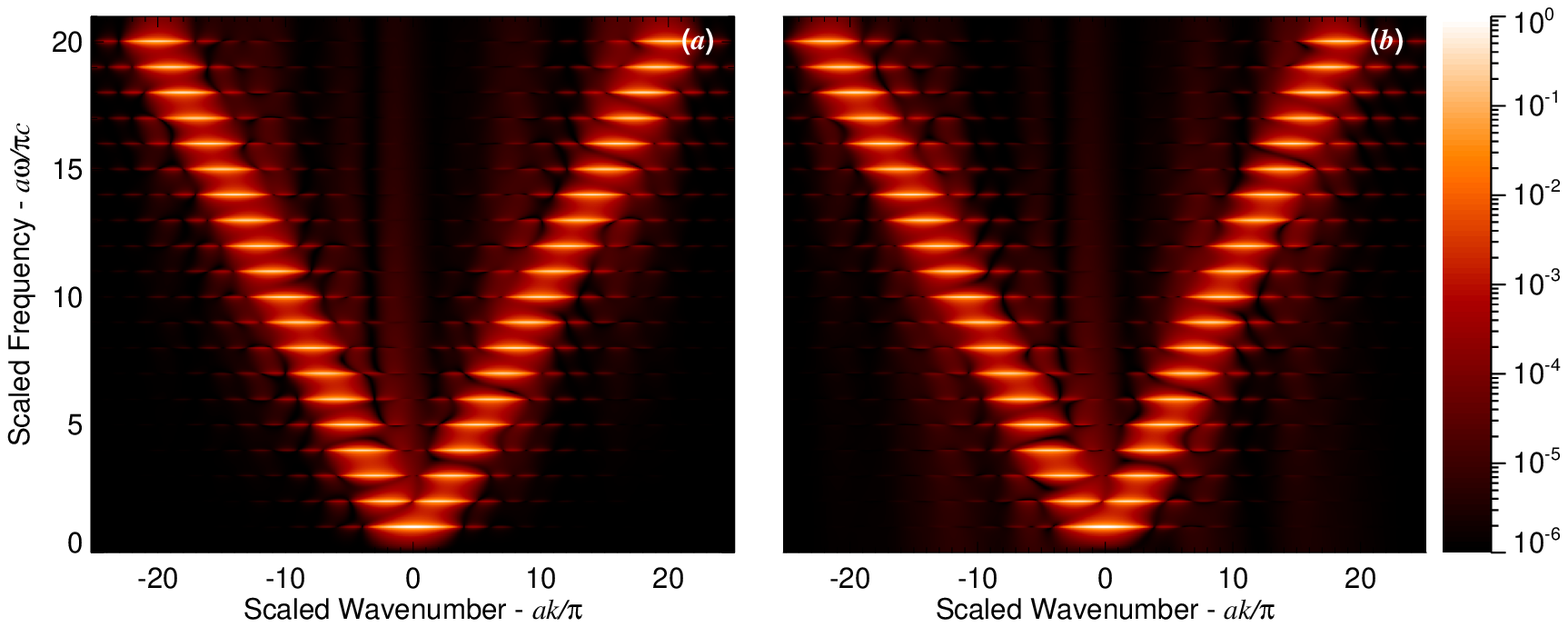}%
        \caption{\small Spectra of waves obtained for the simple 1-D analog
model of meridional circulation. Spectra of the waveform~\eqnref{eqn:RandWalkDamped}
are shown as a function of a dimensionless frequency $a\omega/\pi c$ and
wavenumber $ak/\pi$. The scaling is such that power from the mode with
degree $l$ is concentrated near the points $(\pm l, l)$. ($a$) The power
spectrum in the absence of flow. ($b$) The power spectrum resulting from
a flow with a maximum value of the Mach number $U_0/c = 10^{-1}$ (chosen
to be large for illustrative purposes). The meridional flow varies spatially
as the fundamental sinusoid~\eqnref{eqn:Usine} that vanishes at both ends.
The flow distorts the wavenumber sidelobes arising from spatial windowing.
For both spectra, the damping rate is given by $\eta_l = 2.5 \times 10^{-2} \Delta \omega$,
where $\Delta\omega = \pi c/a$ is the frequency spacing between modes. For
a sound speed appropriate for the middle of the solar convection zone
($c \approx 0.1$ Mm s$^{-1}$) and a domain length of half the circumference of the Sun
at that depth ($a = 600\pi$ Mm), this damping rate is equivalent to a
linewidth $\Gamma = \eta_l/\pi = 1.3 \mu$Hz.
\label{fig:PowSpec0.1} }%

\end{figure*}%
}


\def\figtwo{%
\begin{figure*}%
        \epsscale{0.5}%
        \plotone{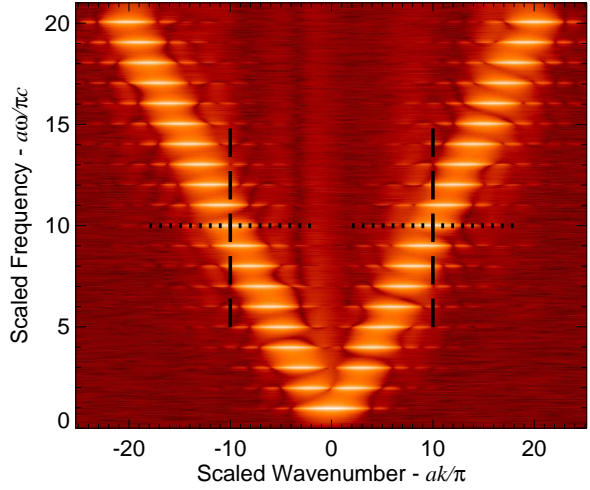}%
        \caption{\small Same as Figure~\ref{fig:PowSpec0.1}$b$, except that
the maximum Mach number is $10^{-3}$ and white Gaussian noise has been added.
The ratio of signal to noise is appropriate for low-degree $p$ modes and
has been estimated from observations from a variety of sources (see \S\ref{subsec:Noise}).
The Doppler shift is difficult to see by eye for such a low Mach number,
and the spatial sidelobes, so prominent in Figure~\ref{fig:PowSpec0.1}, have
been partially obscured. The dashed lines mark the cuts at constant wavenumber
$ak/\pi = \pm l$ shown in Figure~\ref{fig:Cuts}$a$, the dotted lines the cuts
at constant frequency $a\omega/\pi c = l$ shown in \ref{fig:Cuts}$b$.
\label{fig:NoisySpec0.001} }%

\end{figure*}%
}


\def\figthree{%
\begin{figure*}%
        \epsscale{1.0}%
        \plotone{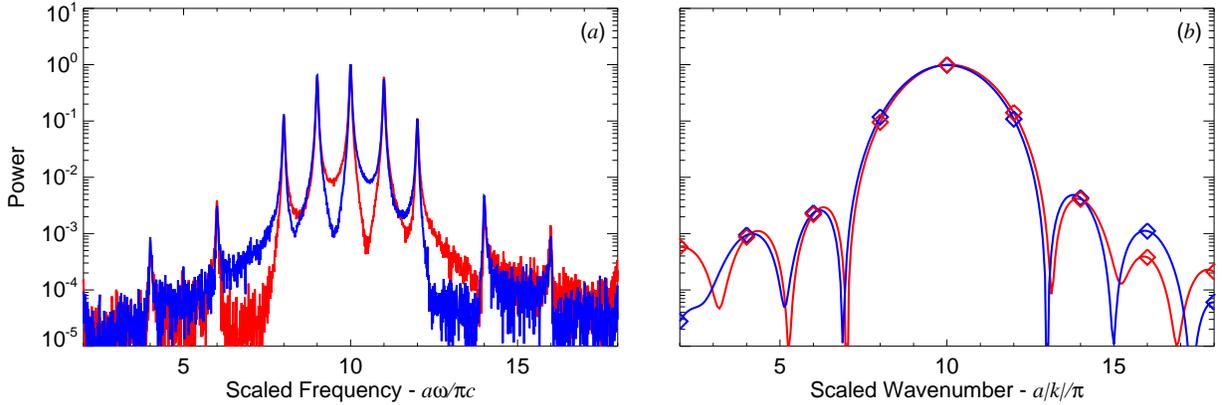}%
        \caption{\small Cuts through the spectrum shown in Figure~\ref{fig:NoisySpec0.001}
at ($a$) constant wavenumber $ak/\pi = \pm l$ and ($b$) constant frequency
$a\omega/\pi c = l$ for $l = 10$. The blue and red curves correspond to waves
propagating in the positive and negative $x$-direction---the right (prograde)
and left (retrograde) branches in Figure~\eqnref{fig:NoisySpec0.001} respectively.
The Doppler effect manifests as a clear shift of the power distribution in the
wavenumber, but not in frequency. In panel ($a$), the spatial sidelobes from
modes with nearby $l$ generate a comb of frequencies that do not shift when a
meridional circulation is present. The amplitude of each individual peak depends
on the spatial windowing, the lifetime of the mode, and weakly on the flow speed
and profile. In panel ($b$), the peaks are the spatial leaks generated by the
$l = 10$ mode, shifted slightly by the background flow.
\label{fig:Cuts} }%

\end{figure*}%
}


\def\figfour{%
\begin{figure*}%
        \epsscale{0.5}%
        \plotone{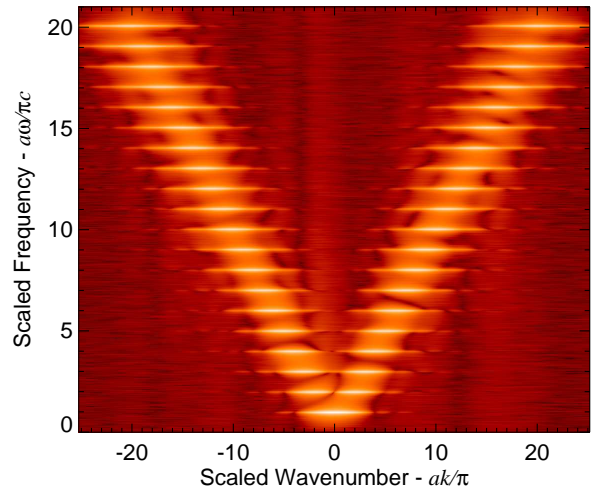}%
        \caption{\small Power spectrum generated for a fast flow ($U_0/c = 0.1$)
with the cubic flow profile given by Equation~\eqnref{eqn:UCubic}. Since
the flow profile is antisymmetric, instead of undergoing a Doppler shift, each
branch is Doppler broadened. The broadening has caused partial merger of
the spatial sidelobes.
\label{fig:CubicFlowPow} }%

\end{figure*}%
}


\def\figfive{%
\begin{figure*}%
        \epsscale{1.0}%
        \plotone{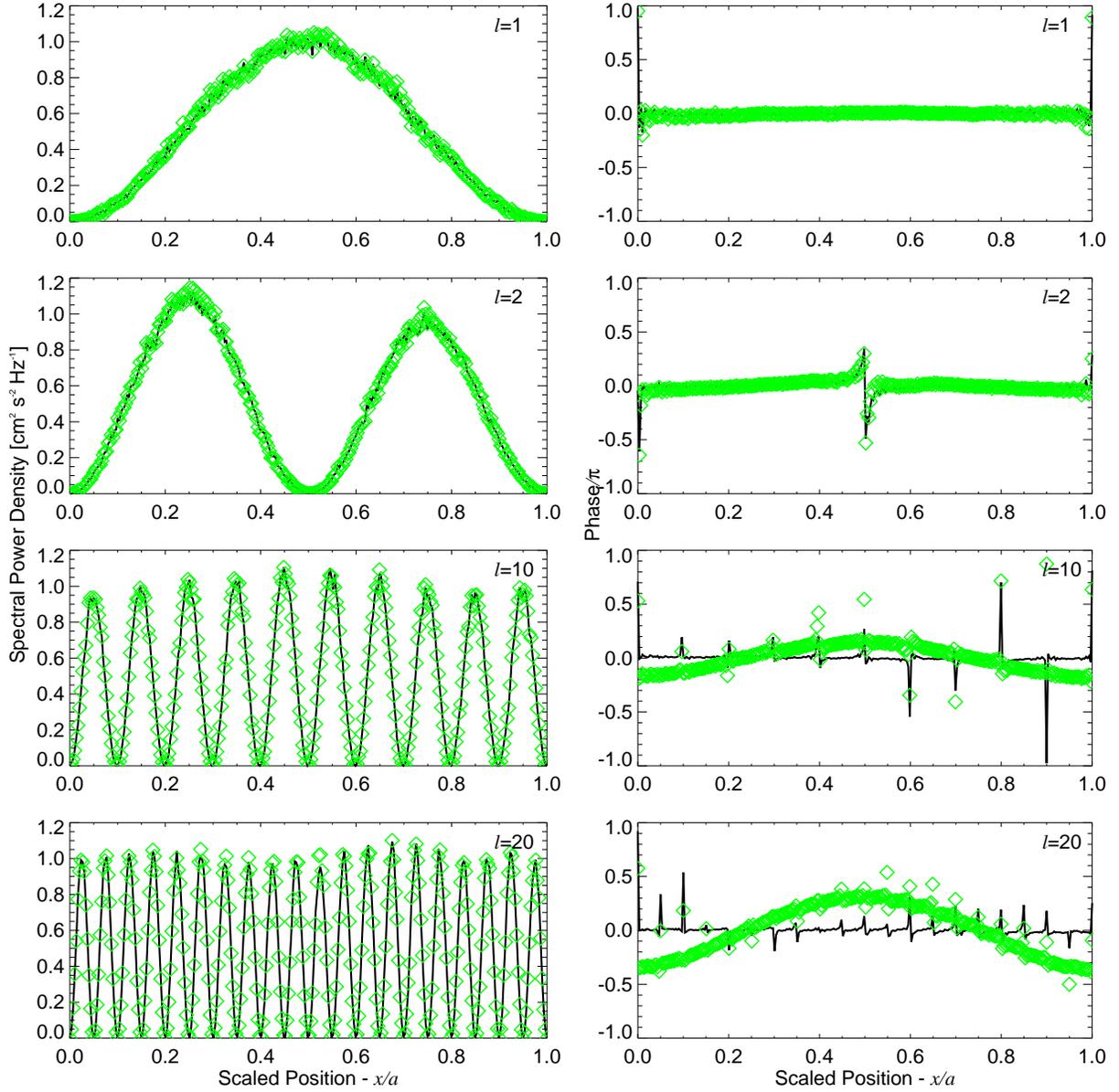}%
        \caption{\small Spectral power density (left column) and phases (right column)
of a subset of modes calculated using Equations~\eqnref{eqn:AveEigfunc}--\eqnref{eqn:Amplitude}
and the artificial data displayed in Figure~\ref{fig:CubicFlowPow}. The origin
of the phase is defined to be zero in the absence of noise and background flow $U$,
and is determined by demanding that the spatial average of $\phi_l$ vanishes. The black
curve shows the results in the absence of flow, and the green diamonds show
the results when the flow field is present. The magnitude of the phase
variation increases linearly with $l$, as expected from Equation~\eqnref{eqn:dphidx}.
\label{fig:PhaseAmps} }%

\end{figure*}%
}


\def\figsix{%
\begin{figure*}%
        \epsscale{0.5}%
        \plotone{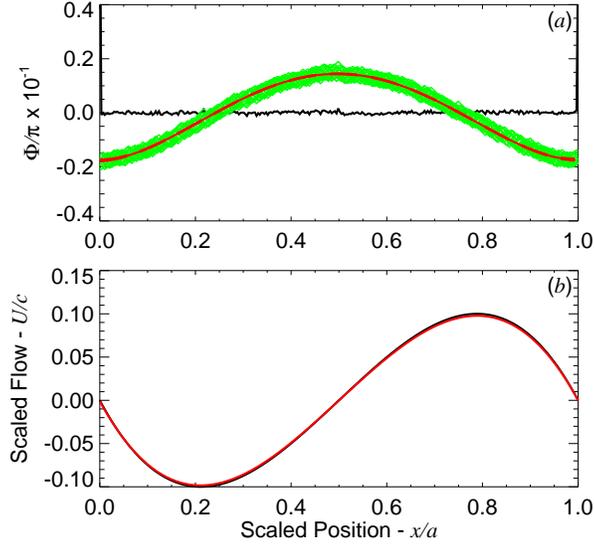}%
        \caption{\small The mean phase function $\Phi(x)$ and the meridional flow
profile derived from the phase function. The flow possesses an antisymmetric, cubic
profile with a maximum Mach number of $U_0/c = 10^{-1}$. ($a$) The mean phase function
$\Phi$ as a function of position $x$. The black curve shows the phase function
in the absence of flow, while the green diamonds (appearing as a blurred band solely
because of the high density of points) show the result in the presence of the flow.
The solid red curve is a low-degree polynomial fit to the phase function.
The errors in the fit are exceedingly small, and thus, even though dashed red curves
have been included to mark $3\sigma$ errors, the error curves are indistinguishable
from the fit itself. ($b$) The input flow field, defined by Equation~\eqnref{eqn:UCubic},
is indicated by the black curve. The flow field derived by differentiating the
red curve in panel $a$ is shown in red (lying nearly on top of the black curve).
For the fast flow considered here, the phase-extraction technique recovers the
input velocity remarkably well.
\label{fig:AvePhase0.1} }%

\end{figure*}%
}


\def\figseven{%
\begin{figure*}%
        \epsscale{0.5}%
        \plotone{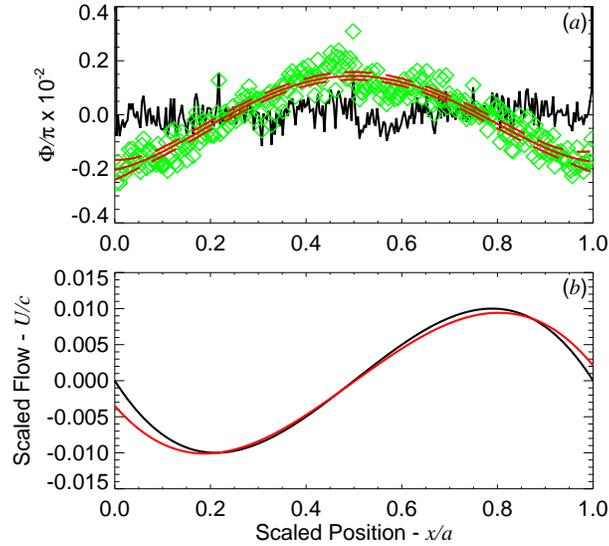}%
        \caption{\small Same as Figure~\ref{fig:AvePhase0.1} except that the
maximum Mach number is smaller: $U_0/c = 10^{-2}$. In panel ($a$) the dashed red
curves mark $\pm3\sigma$ errors in the fitted quantity.
\label{fig:AvePhase0.01} }%

\end{figure*}%
}


\def\figeight{%
\begin{figure*}%
        \epsscale{0.5}%
        \plotone{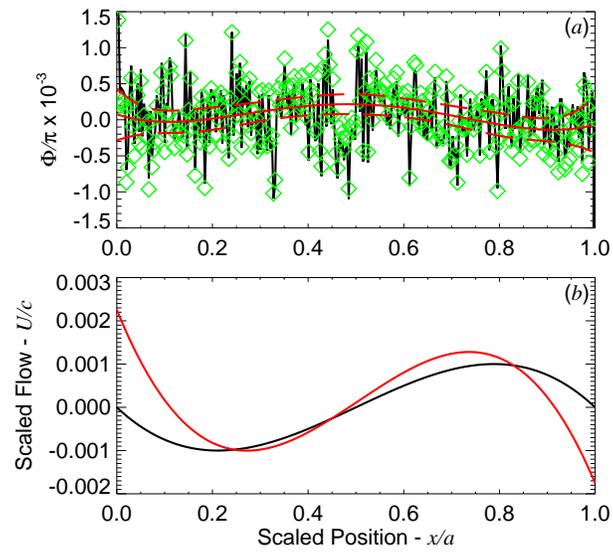}%
        \caption{\small Same as Figure~\ref{fig:AvePhase0.1} except that $U_0/c = 10^{-3}$.
The phase extraction procedure works well in the interior, but has begun to fail at
the edges of the domain where all the eigenfunctions have nodes.
\label{fig:AvePhase0.001} }%

\end{figure*}%
}


\def\fignine{%
\begin{figure*}%
        \epsscale{1.0}%
        \plotone{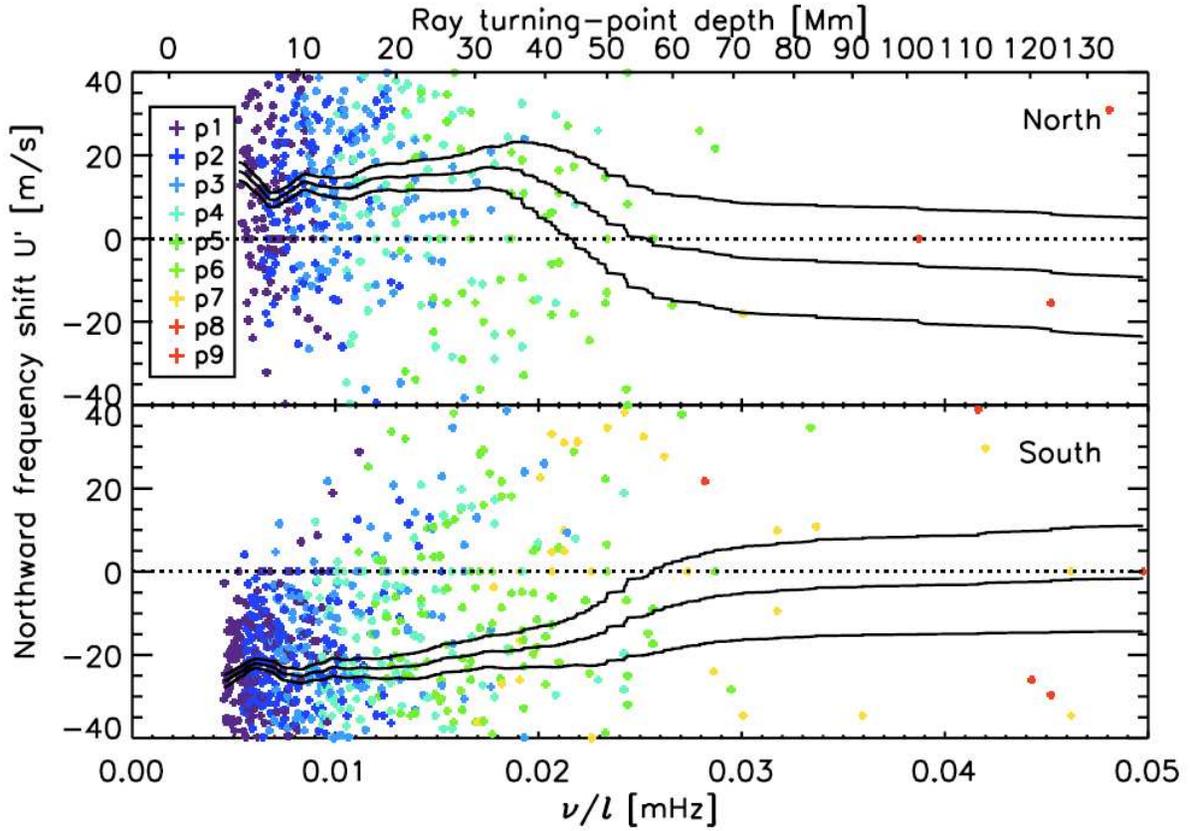}%
        \caption{\small The observed frequency shift deduced from 3 months of
MDI data obtained in 1997. The Doppler shift was measured by cross-correlating the
two power spectra obtained by projecting separately onto northward- and
southward-propagating waveforms. The apparent vanishing of the frequency shift for
waves with a lower turning point greater than 40 Mm may be an artifact caused
by the transition in the spectra from continuous ridges to discrete, normal mode
peaks \citep[after][]{Mitra-Kraev:2007}.
\label{fig:MitraKraev} }%

\end{figure*}%
}

\figone
\figtwo
\figthree
\figfour
\figfive
\figsix
\figseven
\figeight
\fignine

\end{document}